\documentclass[aps,prl,reprint,superscriptaddress,amssymb,amsmath]{revtex4-1}
\usepackage{graphicx}
\usepackage[utf8]{inputenc} 
\usepackage[T1]{fontenc}
\usepackage[USenglish]{babel} 
\usepackage{hyperref}
\hypersetup{
    colorlinks=true,        
    linkcolor=red,          
    citecolor=blue,        
    filecolor=magenta,      
    urlcolor=blue           
}
\usepackage[capitalize]{cleveref}

\begin{document}

\title{Experimental investigation of the optical spin selection rules in bulk Si and Ge/Si quantum dots}

\author{N. Sircar}
  \affiliation{Institut für Experimentelle und Angewandte Physik, Universität Regensburg, D-93040 Regensburg, Germany}
  \affiliation{Walter Schottky Institut, Technische Universität München, Am Coulombwall 4, D-85748 Garching, Germany}
\author{D. Bougeard}
  \affiliation{Institut für Experimentelle und Angewandte Physik, Universität Regensburg, D-93040 Regensburg, Germany}
%
\begin{abstract}
We study the relationship between the circular polarization of photoluminescence and the magnetic field-induced spin-polarization of the recombining charge carriers in bulk Si and Ge/Si quantum dots. First, we quantitatively compare experimental results on the degree of circular polarization of photons resulting from phonon-assisted radiative transitions in intrinsic and doped bulk Si with calculations which we adapt from recently predicted spin-dependent phonon-assisted transition probabilities in Si. The excellent agreement of our experiments and calculations quantitatively verifies these spin-dependent transition probabilities and extends their validity to weak magnetic fields. Such magnetic fields can induce a luminescence polarization of up to 3\%/T. We then investigate phononless transitions in Ge/Si quantum dots as well as in degenerately doped Si. Our experiments systematically show that the sign of the degree of circular polarization of luminescence resulting from phononless transitions is opposite to the one associated with phonon-assisted transitions in Si and with phononless transitions in direct band gap semiconductors. This observation implies qualitatively different spin-dependent selection rules for phononless transitions, which seem to be related to the confined character of the electron wave function.
\end{abstract}

\maketitle

Silicon is an attractive materials platform for spin-based information processing devices \cite{Jansen2012} due to properties favoring long spin lifetimes such as a weak hyperfine coupling \cite{Wild2012}, low spin-orbit coupling and the absence of piezo-electricity \cite{Zutic2004, Fabian2007}. For direct bandgap semiconductors the optical orientation of charge carrier spins by an interaction with circularly polarized light represents an important tool for the study of carrier spins \cite{Meier1984,Dyakonov2008}. For indirect bandgap group IV materials, this concept has triggered recent work, highlighting for example the accessibility of optical spin orientation via the direct bandgap in Ge \cite{Pezzoli2012,Yasutake2013}. In bulk Si, however, optical transitions relevant for spin orientation experiments involve phonon-assisted transitions across the indirect bandgap. A quantitative and contact-free optical spin detection and analysis of spin dynamics in this material will depend on the knowledge of spin-dependent optical selection rules for the involved phonon-assisted transitions. The foundations of such selection rules have been discussed only recently \cite{Li2010,Cheng2011}, providing a theoretical framework. Moreover, the spin-dependent mechanisms governing phononless transitions in Si-based quantum confined structures, which have been discussed in terms of enhanced optical properties compared to bulk Si \cite{Zabel2013, Tsybeskov2009a}, have yet to be established. 

In this contribution, we  present a study on spin-dependent transition probabilities for radiative recombinations of photoexcited carriers in bulk Si and quantum confined Ge/Si structures in photoluminescence (PL) experiments. The degree of spin polarization (DSP) of the photoexcited carriers is adjusted through static magnetic fields. In order to quantitatively connect the DSP with the degree of circular polarization of PL (DCP), we adapt the theory on the DCP of GaAs-type semiconductors in a weak magnetic field \cite{Dyakonov1972} to Si, using recent theoretical predictions on the DCP in Si \cite{Li2010}. Our theoretical description is in excellent quantitative agreement with our measurements of the DCP for phonon-assisted transitions both in intrinsic and degenerately doped Si. A substantial DCP of up to 3\%/T at liquid He temperatures highlights the impact of an applied magnetic field in optical orientation experiments in Si. Regarding the spin-dependent selection rules of phononless transitions in indirect bandgap semiconductors, such as those occurring in Ge/Si quantum dots (QDs), our comparative study of experiments and calculations suggests that band-mixing effects due to the quantum confinement play a significant role.

In the following, we use the definition $ \text{DCP} \equiv (I_{\sigma^+}-I_{\sigma^-})/(I_{\sigma^+}+I_{\sigma^-})$, where $I_{\sigma^+}$ and $I_{\sigma^-}$ are the intensities of $\sigma^+$- and $\sigma^-$-polarized components of the PL, respectively. Furthermore, the spin quantization axis and hereby the meaning of $\sigma^\pm$ are always defined by the positive direction of the external magnetic field. Under weak-field conditions, the magnetic field-induced DCP for semiconductors of GaAs type can be separated into two additive contributions of polarized electrons, $\text{DCP}^\text{e}$, on the one hand and holes, $\text{DCP}^\text{h}$, on the other hand, which recombine with unpolarized holes and electrons, respectively. In a first-order approximation, \cite{Dyakonov1972}
\begin{equation}
\text{DCP} = \text{DCP}^\text{e} + \text{DCP}^\text{h} = -\left\langle s_\text{z}\right\rangle +\left\langle j_\text{z}\right\rangle. 
\label{eq:PL_polarization_field_holes_and_electrons}
\end{equation}
$\left\langle s_\text{z}\right\rangle$ and $\left\langle j_\text{z}\right\rangle$ are the average projections of the electron and hole angular quantum numbers along the field direction. 

To account for the different spin-dependent transition probabilities in Si as compared to GaAs-type semiconductors, we modify \cref{eq:PL_polarization_field_holes_and_electrons} to
\begin{equation}
\text{DCP}_m = 2\eta_m \left\langle s_\text{z}\right\rangle + \frac{4}{3} \gamma_m \left\langle j_\text{z}\right\rangle. 
\label{eq:PL_polarization_field_holes_and_electrons_Si}
\end{equation}
Here $m$ denotes either the transverse-acoustic (TA), transverse-optical (TO), longitudinal-acoustic (LA) or longitudinal-optical (LO) phonon-mode involved in the considered radiative recombination. $\eta_m$ and $\gamma_m$ are the maximally attainable PL polarizations when electrons and holes are fully oriented in the $s_\text{z}=+1/2$ and $j_\text{z}=+3/2$ states, respectively. They contain the information on the spin-dependent selection rules. We have introduced the numerical prefactors 2 and 4/3 to ensure that \cref{eq:PL_polarization_field_holes_and_electrons_Si} equals \cref{eq:PL_polarization_field_holes_and_electrons} if $\eta_m$ and $\gamma_m$ were deduced from the spin-dependent transition probabilities of direct transitions for GaAs-type semiconductors. As $\left\langle s_\text{z}\right\rangle$ and $\left\langle j_\text{z}\right\rangle$ are well-known functions of the applied magnetic field, the spin-dependent selection rules of phonon-assisted radiative transitions in Si can thus be assessed by measuring $\text{DCP}_m$.

While $\eta_m$ has been explicitly calculated in Ref.~\citep{Li2010}, we derive the values of $\gamma_m$ from the strict symmetry arguments outlined in the same reference: The intensity of the $\sigma^+$-polarized component of luminescence arising from the recombinations of holes originating from the $j_\text{z}=+3/2$ hole states is  $\smash{I^{\text{h},x\vphantom{y}}_{\sigma^+,m} = I^{\text{h},y}_{\sigma^+,m}= \frac{9}{16} I_{0,m}^{x,y}}$ for the transitions ending in the $x$- and $y$-valleys, which have their axis of revolution perpendicular to the spin quantization axis. The intensity of the of $\sigma^-$-polarized light is $\smash{I^{\text{h},x\vphantom{y}}_{\sigma^-,m} = I^{\text{h},y}_{\sigma^-,m}= \frac{1}{16} I_{0,m}^{x,y}}$ for each of these four transitions. Here, $I_{0,m}^{x,y}$ denotes the total luminescence intensity due to all transitions assisted by transverse phonons with final states in the $x$- and $y$-valleys, respectively. Transitions ending in the two $z$-valleys do not originate from the $+3/2$ state, and can thus be disregarded in the context of polarization. Nevertheless, these latter transitions will contribute to the overall luminescence with an intensity $\smash[t]{I_{0,m}^{z\vphantom{y}}}$. We thus end up with $
\gamma_{\text{TO}}    = \frac{2I^{\text{h},x\vphantom{y}}_{\sigma^+,\text{TO}}+ 2I^{\text{h},y}_{\sigma^+,\text{TO}} - 2I^{\text{h},x\vphantom{y}}_{\sigma^-,\text{TO}} - 2I^{\text{h},y}_{\sigma^-,\text{TO}}}{4 I_{0,\text{TO}}^{x,y}+2 I_{0,\text{TO}}^{z\vphantom{y}}}
										 =0.293$ and likewise with $\gamma_{\text{TA}} = 0.325$, where we employed the numerical values of $\smash[t]{I^{x,y}_{0,m}}$  and $\smash[t]{I_{0,m}^{z\vphantom{y}}}$ as given in Ref. \citep{Li2010}. A similar analysis of the selection rules for the LO-assisted recombinations yields $\gamma_{\text{LO}} = - 0.375$.

In our experiments, the PL was excited with either $\lambda_\text{exc}$= 532 nm or 514 nm light with linear polarization. The samples were mounted in a variable temperature magnet cryostat and the PL recorded in Faraday geometry defined by the direction of the external magnetic field. We first present measurements on bulk Si samples, cut from commercial, single-crystalline, high-quality Si wafers. High-purity Si with a residual impurity concentration below $3\times 10^{12}\text{cm}^{-3}$, referred to as intrinsic Si, degenerately Sb-doped $n$-type Si ($N_\text{D}=8\times 10^{18}\text{cm}^{-3}$) and degenerately B-doped $p$-type Si ($N_\text{A}=2.3\times 10^{19}\text{cm}^{-3}$) were studied. We also discuss experiments on Ge QD structures fabricated with solid-source molecular beam epitaxy, which contained 80 QD layers in total, each separated by 25 nm-thick Si layers. Details on the growth, morphology, composition, and electronic band structure of this specific QD sample may be found elsewhere.\cite{Tan2003,Bougeard2004}

\begin{figure}
\centering
\includegraphics[width=\columnwidth]{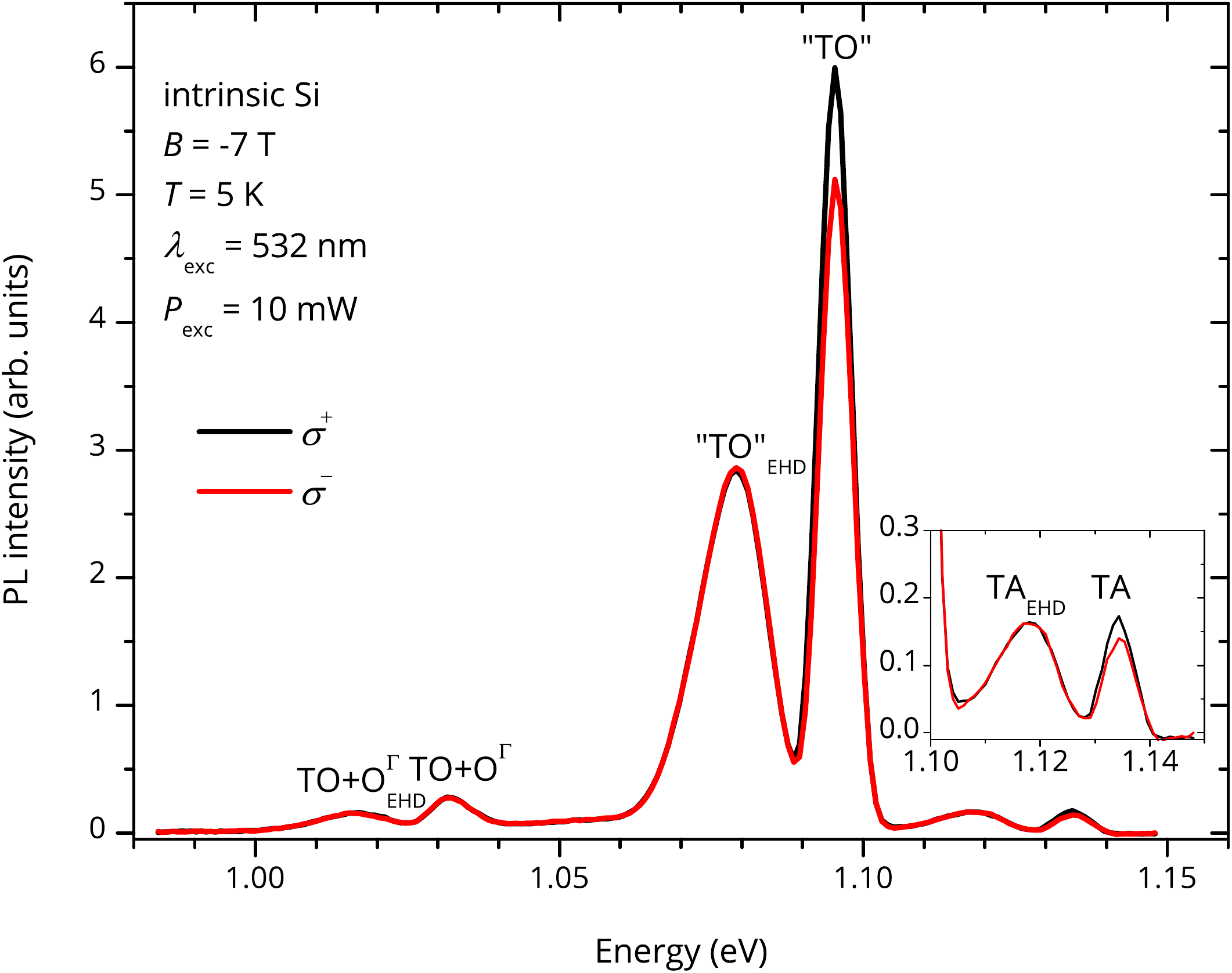}
\caption{$\mathnormal{\sigma}^+$- and $\mathnormal{\sigma}^-$-polarized PL spectra of the intrinsic Si sample in a magnetic field of -7 T. The inset is a close-up for the TA-related luminescence. Both the ``TO'' and TA free-exciton peaks exhibit a positive DCP. See the text for a definition of ``TO''.}
\label{fig:Si_n18_PL_vs_energy_-7T_polarization}
\end{figure}
\Cref{fig:Si_n18_PL_vs_energy_-7T_polarization} depicts the spectra of the $\sigma^+$- and $\sigma^-$-polarized components of the PL of the intrinsic Si sample in a field of -7 T and at a temperature of 5 K. In these spectra, one can distinguish three well-resolved main luminescence peaks each accompanied by a broader side peak. The main peaks correspond to the TA, ``TO'', and $\text{TO+O}^\Gamma$ phonon replica of free-exciton recombinations as indicated in the figure \cite{Dean1967}. In fact, for the present measurement the ``TO'' peak is a mixture of a TO and a weaker LO component \cite{Davies1989}. As \cref{fig:Si_n18_PL_vs_energy_-7T_polarization} reveals, both the ``TO'' free-exciton peak, as well as the TA feature, clearly display a positive DCP.  

\begin{figure}
\centering
\includegraphics[width=\columnwidth]{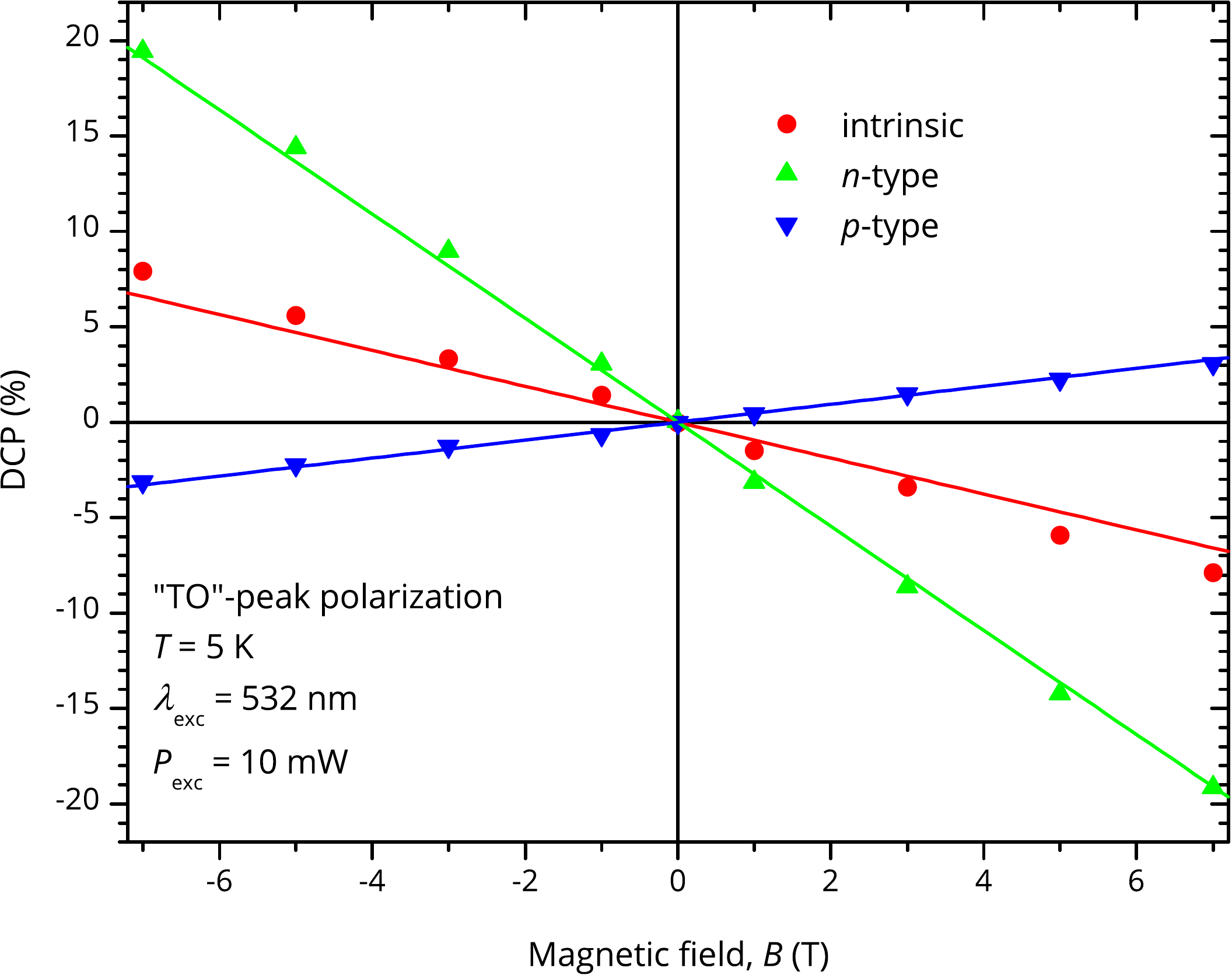}
\caption{Magnetic field dependence of the DCP of the ``TO'' PL peak of bulk Si of different doping type. The symbols give experimental values while the solid lines represent the theoretical calculations presented in the text.}
\label{fig:Si_polarization_B_sweep}
\end{figure}

In the following, we focus on the ``TO'' replica peak in more detail. $\text{DCP}_\text{``TO''}$ was measured for the ``TO'' peak energy for different external magnetic fields up to 7 T at a temperature of 5 K. The resulting field dependences of $\text{DCP}_\text{``TO''}$ for the intrinsic and the heavily doped $p$-type and $n$-type samples are plotted in \cref{fig:Si_polarization_B_sweep}. Within the employed field range, $\text{DCP}_\text{``TO''}$ is a linear function of the field for all samples. Also, as expected, it vanishes for zero-field. Interestingly, while the slope of the field dependence is negative for both the intrinsic and the $n$-type sample, it is positive for the $p$-type sample. This behavior indicates that the dominant polarization mechanism of the luminescence is qualitatively different in these two cases.

For intrinsic Si, Boltzmann statistics describe both the electron and hole system. Hence, $\left\langle s_\text{z}\right\rangle$ and $\left\langle j_\text{z}\right\rangle$ of \cref{eq:PL_polarization_field_holes_and_electrons_Si} are given in first order approximation as $\left\langle s_\text{z}\right\rangle \approx - \frac{1}{4} \frac{g_\text{e} \mu_\text{B} B}{k_\text{B} T}$ and $\left\langle j_\text{z}\right\rangle \approx - \frac{5}{4} \frac{g_\text{h} \mu_\text{B} B}{k_\text{B} T}$ with the applied magnetic field $B$, the temperature $T$, the Boltzmann constant $k_\text{B}$, the Bohr magneton $\mu_\text{B}$, the electron $g$-factor $g_\text{e}$, and the hole $g$-factor $g_\text{h}$ \cite{Dyakonov1972}. We use the values $g_\text{e}=2$ and $g_\text{h}=0.56$ \cite{Feher1959,Suzuki1964}. To account for the admixture of the  LO-component to the ``TO'' peak we calculate the effective polarization factor $\eta_{\text{``TO''}}$ by a weighted averaging over the contributing polarization factors $\eta_{\text{TO}}$ and $\eta_{\text{LO}}$. The relative intensities $I_{0,\text{TO}}$ and $I_{0,\text{LO}}$ of the TO and LO peak \cite{Li2010}, respectively, are taken as weighting factors. This results in $ \eta_{\text{``TO''}}  = \frac{\eta_{\text{TO}} I_{0,\text{TO}} + \eta_{\text{LO}} I_{0,\text{LO}}}{I_{0,\text{TO}}+I_{0,\text{LO}}} = -0.133$. Using the $\gamma_{\text{TO}}$ and $\gamma_{\text{LO}}$ values that we have calculated earlier, an effective polarization factor $\gamma_{\text{``TO''}}=0.285$, attributed to polarized holes, is obtained by a similar weighting procedure. Thus according to \cref{eq:PL_polarization_field_holes_and_electrons_Si}, the general expression for the field and temperature dependence of $\text{DCP}_\text{``TO''}$ of intrinsic Si is then $\text{DCP}_\text{``TO''} = -0.067 \cdot \frac{\mu_\text{B} B}{k_\text{B} T}$. A straight line representing this latter expression for a temperature of $T=5\text{ K}$ is shown in \cref{fig:Si_polarization_B_sweep}. The experimental data points for intrinisc Si are in excellent agreement with our theoretical prediction. The experimental observation and theoretical finding of a negative sign of the slope of $\text{DCP}_\text{``TO''}$ for intrinsic Si congruently hint toward the importance of the polarization of the hole angular momentum in a magnetic field, which dominates the PL polarization.

Compared to intrinsic Si, the slope of heavily $n$-type-doped Si is also negative, but steeper. This can be connected to the degenerate character of the electron system, which now obeys the Fermi-Dirac statistics. The average spin moment of the electron system in this case is thus given by $\left\langle s_\text{z}\right\rangle = -\frac{3}{8} \frac{g_\text{e} \mu_\text{B} B}{E_\text{F}^\text{e}}$ \cite{Dyakonov1972}. For the holes, on the other side, Boltzmann statistics still apply, so that the expression for their angular momentum does not differ from the one of intrinsic Si. According to \cref{eq:PL_polarization_field_holes_and_electrons_Si}, the PL polarization in a magnetic field becomes here $\text{DCP}_\text{``TO''} =- \frac{3}{4} \frac{g_\text{e} \mu_\text{B} B}{E_\text{F}^\text{e}} \eta_{\text{``TO''}} - \frac{5}{3}  \frac{g_\text{h}\mu_\text{B} B}{k_\text{B} T} \gamma_{\text{``TO''}}$. At the studied doping concentration of about $N_\text{D}=8\times 10^{18}\text{ cm}^{-3}$, reasonable values for the electron Fermi energy lie around $E_\text{F}^\text{e} \approx 20 \text{ meV}$ \cite{Dumke1983}. Hence, the first term in the summation of $\text{DCP}_\text{``TO''}$ is negligible. For degenerate $n$-type Si the polarization of PL is therefore only a consequence of the polarization of the excited holes. Under this premise, and taking the same value for $\gamma_{\text{``TO''}}$ as for intrinsic Si, the field-dependence of this expression of $\text{DCP}_\text{``TO''}$ is also plotted in \cref{fig:Si_polarization_B_sweep}. As can be seen, the calculated curve perfectly describes the measured field-dependence of $n$-type doped Si.

For $p$-type Si the situation is reversed compared to the $n$-type case. The hole angular momentum  $\left\langle j_\text{z}\right\rangle$ can now be neglected due to considerations similar to those for the $n$-type doping, but now for the Fermi energy of the hole gas \cite{Dumke1983a}. $\left\langle s_\text{z}\right\rangle$ remains the same as in the intrinsic case, i.e. $\text{DCP}_\text{``TO''} = - \frac{1}{2}  \frac{g_\text{e}\mu_\text{B} B}{k_\text{B} T} \eta_{\text{``TO''}}$. Yet, we have to take into account that for a high $p$-type doping, spin-orbit coupling comes into effect as a consequence of the proximity of the Fermi energy to the split-off band \cite{Li2010}. Given the doping concentration of about $N_\text{A}=2.3\times 10^{19}\text{cm}^{-3}$, the polarization factor $\eta_{\text{``TO''}}$ then decreases to $\eta_{\text{``TO''}} =  -0.035$ \cite{Li2010}. We included the straight line representing the field dependence of the corresponding $\text{DCP}_\text{``TO''}$ in \cref{fig:Si_polarization_B_sweep}. Again, the matching of the experiment and theory is excellent. In this case, the slope is now positive, revealing a dominant contribution of the recombination of spin-polarized electrons to the luminescence polarization. 

\begin{figure*}
\centering
\includegraphics[width=0.75\textwidth]{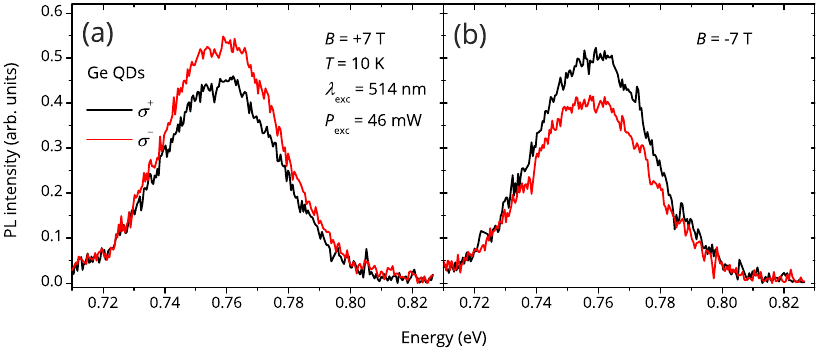}
\caption{Circular-polarization-resolved spectra of the phononless PL of the Ge QD sample. The spectra were acquired at a temperature of about 10 K and in a magnetic field of (a) +7 T and (b) -7~T.}
\label{fig:Ge_hut_cluster_PL_pol_B=7T}
\end{figure*}

For each studied case --- intrinsic, degenerate $n$-type and degenerate $p$-type Si --- our respective calculations, which are based on selection rules presented in \cite{Li2010}, show an excellent quantitative agreement with the experimentally observed magnetic field dependence of the DCP. Our measurements represent the first direct and quantitative experimental confirmation of the spin-dependent transition probabilities from Ref.  \cite{Li2010}. Additionally, we show here that the validity of these transition probabilities can be extended to weak magnetic fields. The consistency between experiment and theory in the presence of a weak external field suggests that merely the band population depending on the angular momentum is changed by the field, whereas the matrix elements of the dipole and the phonon-assisted transitions between the individual bands, which mainly rely only on the band symmetries, are not affected. The observed DCP of up to 3\%/T represents a significant and entirely field-induced contribution to the luminescence polarization of optical spin injection or detection experiments in Si when carried out in the presence of an external magnetic field, such as for example in spin LEDs \cite{Jonker2007,Kioseoglou2009, Li2009, Jansen2010a}.

Our experiments show that doping plays a critical role in the magnitude of the field-induced polarization effect. In the cases of strong $p$- and $n$-type degeneracy we could completely disregard respectively either holes or electrons in the analysis of the data. In this context, it is interesting to highlight the broad side-peaks, each approximately 17 meV lower in energy than the corresponding phonon-replica peak, as shown in \cref{fig:Si_n18_PL_vs_energy_-7T_polarization}. These are due to the recombination of electron-hole pairs within electron-hole droplets (EHD) at low temperatures \cite{Keldysh1986}. As can be seen, these features do not exhibit any signs of a circular polarization up to fields of 7 T. Since the quasi-Fermi energies of both the electron and hole system are relatively large in the EHD \cite{Vashishta1974}, this behavior is fully in line with our previous description of degenerate systems: i.e. now in the EHD, both carrier species do not exhibit a relevant angular momentum polarization and no DCP is observed in the experiment.

Having discussed the spin-dependent phonon-assisted selection rules in Si, we now address spin-dependent phononless radiative recombinations. These occur in indirect band-gap materials whenever momentum mismatch in optical transitions may be removed, e.g. by quantum confinement. We first present the PL spectra for the Ge QD sample which are dominated by phononless radiative recombinations \cite{Brunner2002}. 

\Cref{fig:Ge_hut_cluster_PL_pol_B=7T}a and \ref{fig:Ge_hut_cluster_PL_pol_B=7T}b depict the polarization-resolved phononless PL of the Ge QD sample in a magnetic field of +7 and -7 T, respectively. They clearly exhibit a circular polarization, the field dependence of the DCP having a negative gradient. 

For these undoped Ge QDs neither the spin polarization of electrons, which are localized in Si, nor that of holes, which are localized in Ge \cite{Bougeard2004}, can be neglected. Recombination of both species thus contribute to the observed polarization of luminescence. Since the electrons are weakly confined in Si at the interface to the Ge QD, the magnetic field dependence of the average spin momentum  $\left\langle s_\text{z}\right\rangle$ of these electrons is similar to electrons in bulk Si. However, the orientation of the angular momentum $\left\langle j_\text{z}\right\rangle$ of the holes confined within the Ge is reversed from the bulk Si case, because the holes now possess a negative hole $g$-factor \footnote{$g_{\text{h}}$ in bulk Ge is negative \cite{Stickler1962, Hensel1969}. Given the compositional and structural parameters of the present QDs, the negative sign is affected neither by the slight alloying of Ge with Si \cite{Fraj2007} nor by the quantum confinement of the holes\cite{Nenashev2003a}. Experiments on larger GeSi islands do also suggest a negative $g_{\text{h}}$ \cite{KatsarosG.2010}}
\nocite{Stickler1962, Hensel1969,Fraj2007,Nenashev2003a,KatsarosG.2010}.

So far, calculations of spin-dependent probabilities of phononless transitions in Ge QDs do not exist. As highlighted by \citet{Fukatsu1997}, phononless PL can be reduced to an effective matrix element of the form $\left\langle \Gamma_8^+\left|\mathbf{\hat{p}}\right|\Delta_1\right\rangle$, where $\mathbf{\hat{p}}$ denotes the dipole operator and $\Gamma_8^+$ and $\Delta_1$ the representations of the hole and electron wave functions, respectively. The $\Gamma_8^+$ valence and $\Delta_1$ conduction band states in the Ge/Si QD system share, at least partly, the same $p$- and $s$-symmetries as, respectively, the valence and conduction band states in GaAs-like semiconductors. This fact motivates us to qualitatively apply the selection rules for GaAs-like radiative recombinations to our Ge QDs in a first approximation. Doing so implies that phononless recombinations would correspond to a negative value for $\eta_m$ and a positive value for $\gamma_m$. Under this premise, a positive slope of the degree of polarization of the Ge QD PL as a function of the external magnetic field is then always expected from \cref{eq:PL_polarization_field_holes_and_electrons_Si}, regardless of the exact values of $\eta_m$, $\gamma_m$, $ \left\langle s_\text{z}\right\rangle$, and $\left\langle j_\text{z}\right\rangle$. On the contrary, however, the measurements shown in \cref{fig:Ge_hut_cluster_PL_pol_B=7T} yield a negative slope. 

\begin{figure*}
\centering
\includegraphics[width=0.75\textwidth]{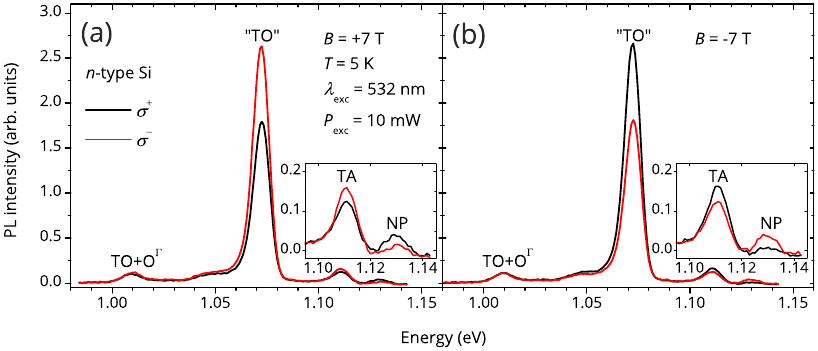}
\caption{Circular-polarization-resolved PL spectra of an $n$-type doped Si sample. The spectra were acquired at a temperature of about 5 K and in a magnetic field of (a) +7 T and  (b) -7 T. The insets are close-ups of the high-energy regions of the spectra, containing the TA peak around 1110 meV and the phononless peak, labeled NP, at 1130 meV. The phononless NP peak has an opposite polarization compared to the TA and ``TO'' peak.}
\label{fig:Si_n3_PL_pol_B=7T_merged}
\end{figure*}
Further evidence for the unexpected behavior of phononless luminescence in indirect band gap materials comes from our PL measurements of doped bulk Si. As a representative example, \Cref{fig:Si_n3_PL_pol_B=7T_merged}a and \ref{fig:Si_n3_PL_pol_B=7T_merged}b depict the polarization-resolved PL spectra of the heavily doped $n$-type Si sample in a magnetic field of +7 T and -7 T, respectively. Due to the high dopant concentration, a weak phononless PL is observed in addition to the phonon-replica as highlighted in the insets of \cref{fig:Si_n3_PL_pol_B=7T_merged}. Interestingly, the phononless luminescence feature at 1130 meV exhibits a sign of the DCP that is opposite to all the transverse-phonon-related luminescence features.
However, if GaAs-like selection rules were applicable for these phononless recombinations, their DCP would have the same sign as the DCP of transverse-phonon-assisted recombinations, according to \cref{eq:PL_polarization_field_holes_and_electrons_Si}, because of the same signs of $\eta_m$ and $\gamma_m$ in GaAs-like and Si transverse-phonon-assisted transitions. 

As a consequence, we thus infer from our experimental observations that the selection rules governing spin-dependent dipole transitions of phononless recombinations in the present samples are different from phonon-assisted recombinations in Si and have to be distinctively different from the ones of GaAs-like direct band gap recombinations. We note that we have observed the exact same behavior for phononless luminescence also from SiGe/Si quantum well samples in PL and electro-luminescence experiments. A possible explanation may lie in the quantum confinement of the electron within the Si material, which, in principle, can lead to an admixture of the $\Gamma^-_6$ states of the Brillouin zone center to the wave function \cite{Fukatsu1997}. It was recently highlighted in a theoretical work that due to the $p$-symmetry of the $\Gamma^-_6$ band in Si the role of $\sigma^-$ and $\sigma^+$ photons for spin-dependent transitions of the form $\left\langle \Gamma^+_8\left|\mathbf{\hat{p}}\right|\Gamma_6^-\right\rangle$ at the direct band gap is inverted in comparison to the same transitions in GaAs \cite{Nastos2007}. Our study demonstrates that it is worthwhile to include such band mixing in a theoretical treatment of spin-dependent selection rules for Si-based quantum confined structures. This would pave the way for quantitative analyses of optical spin detection experiments in quantum-confined structures, such as SiGe quantum wells or Ge/Si quantum dots, and it would extend the understanding of optical spin orientation in silicon.  

\begin{acknowledgments}
The authors would like to thank Jaroslav Fabian for helpful discussions, Gerhard Abstreiter for support and the Deutsche Forschungsgemeinschaft DFG for funding via SPP 1285.
\end{acknowledgments}

\end{document}